\documentclass[journal]{IEEEtran}

\usepackage{tabularx}
\usepackage{array}
\usepackage{cite}
\usepackage{amsmath,amssymb,amsfonts}
\usepackage{algorithmic}
\usepackage{graphicx}
\usepackage{subfigure}
\usepackage{textcomp}
\usepackage{xcolor}
\usepackage{amsmath}
\usepackage{algorithm}
\usepackage{algorithmic}
\usepackage{graphicx}
\usepackage{orcidlink}
\usepackage{acronym}
\usepackage[inline]{enumitem}

\newcommand{\sectionname}{Sect.}
\newcommand{\meter}{\,\mathrm{m}}

\newcommand{\dB}{\,\mathrm{dB}}
\newcommand{\dBm}{\,\mathrm{dBm}}

\newcommand{\MHz}{\,\mathrm{MHz}}
\newcommand{\GHz}{\,\mathrm{GHz}}
\newcommand{\THz}{\,\mathrm{THz}}

\newcommand{\ms}{\,\mathrm{ms}}
\newcommand{\mus}{\,\mu\mathrm{s}}

\newcommand{\mps}{\,\mathrm{m/s}}

\acrodef{AWGN}{additive white Gaussian noise}
\acrodef{AdamW}{adaptive moment estimation with weight decay}
\acrodef{B5G}{beyond-5G}
\acrodef{CSA}{central satellite}
\acrodef{CFmMIMO}{cell-free massive \acs{MIMO}}
\acrodef{mMIMO}{massive \acs{MIMO}}
\acrodef{DPD} {disk Poisson Process} 
\acrodef{DRA}{direct radiating antenna}
\acrodef{DC}{direct current}
\acrodef{DoD}{depth of discharge}
\acrodef{eMBB}{enhanced mobile broadband}
\acrodef{E-GPW}{extended gridded population of world database}
\acrodef{FDM}{frequency division multiplexing}
\acrodef{FIR}{finite impulse response}
\acrodef{TFoA}{thinned \acs{FoA}}
\acrodef{FoA}{formation of arrays}
\acrodef{GEO}{geostationary Earth orbit}
\acrodef{GSO}{geosynchronous Earth orbit}
\acrodef{H-RRM}{heuristic radio resource management}
\acrodef{KPI}{key performance indicator}
\acrodef{IF}{intermediate frequency}
\acrodef{LEO}{low Earth orbit}
\acrodef{LoS}{line-of-sight}
\acrodef{MAI}{multiple access interference}
\acrodef{MB}{multi beam}
\acrodef{MEO}{medium Earth orbit}
\acrodef{mMTC}{massive machine-type communications}
\acrodef{eMTC}{enhanced machine-type communications}
\acrodef{PDD}{Poisson disk distribution}
\acrodef{EIRP}{effective isotropic radiated power}
\acrodef{NTN}{non terrestrial network}
\acrodef{URLLC}{ultra-reliable low-latency communications}
\acrodef{ECEF}{Earth-centered Earth-fixed}
\acrodef{GPS}{global positioning system}
\acrodef{HTFS}{high throughput fractionated satellite}
\acrodef{UHF}{ultra-high frequency}
\acrodef{FF}{formation flying}
\acrodef{MIMO}{multiple input multiple output}
\acrodef{M-MIMO}{massive multiple input multiple output}
\acrodef{GNC}{guidance navigation and control}
\acrodef{GNSS}{global navigation satellite system}
\acrodef{OFDM}{orthogonal frequency division multiplexing}
\acrodef{PHY}{physical-layer}
\acrodef{SA}{sub-array}
\acrodef{SADM}{solar array drive mechanism}
\acrodef{SG}{solar generator}
\acrodef{SNR}{signal-to-noise ratio}
\acrodef{SIR}{signal-to-interference ratio}
\acrodef{SINR}{signal-to-interference-plus-noise ratio}
\acrodef{SSPA}{solid-state power amplifier}
\acrodef{RF}{radio frequency}
\acrodef{R-GEO}{regional \acs{GEO}}
\acrodef{UT}{user terminal}
\acrodef{UE}{user equipment}
\acrodef{QVD}{quantized virtual distancing}
\acrodef{TDM}{time division multiplexing}
\acrodef{R-GEO}{regional GEO}
\acrodef{RRM}{radio resource management}
\acrodef{SSPA}{solid-state power amplifier}
\acrodef{BLER}{block error rate}
\acrodef{DL}{downlink}
\acrodef{UL}{uplink}
\acrodef{PSD}{power spectral density}
\acrodef{UDSM}{ultra deep sub-micron}
\acrodef{PEPS}{power energy platform simulation}
\acrodef{SE}{spectral efficiency}
\acrodef{wrt}{with respect to}
\acrodef{OBP}{on-board digital processor}
\acrodef{UDSM}{ultra-deep-sub-micron}

\acrodef{3GPP}{third generation partnership project}
\acrodef{AWGN}{additive white Gaussian noise}
\acrodef{B5G}{beyond-5G}
\acrodef{CS}{central satellite}
\acrodef{DRA}{direct radiating antenna}
\acrodef{DC}{direct current}
\acrodef{DoD}{depth of discharge}
\acrodef{eMBB}{enhanced mobile broadband}
\acrodef{FDM}{frequency division multiplexing}
\acrodef{FIR}{finite impulse response}
\acrodef{TFoA}{thinned \acs{FoA}}
\acrodef{FoA}{formation of arrays}
\acrodef{GEO}{geostationary Earth orbit}
\acrodef{GSO}{geosynchronous Earth orbit}
\acrodef{KPI}{key performance indicator}
\acrodef{IF}{intermediate frequency}
\acrodef{LEO}{low Earth orbit}
\acrodef{LoS}{line-of-sight}
\acrodef{NLoS}{non-line-of-sight}
\acrodef{MAI}{multiple access interference}
\acrodef{MEO}{medium Earth orbit}
\acrodef{mMTC}{massive machine-type communications}
\acrodef{eMTC}{enhanced machine-type communications}
\acrodef{EIRP}{effective isotropic radiated power}
\acrodef{NTN}{non terrestrial network}
\acrodef{URLLC}{ultra-reliable low-latency communications}
\acrodef{ECEF}{Earth-centered Earth-fixed}
\acrodef{GPS}{global positioning system}
\acrodef{HTFS}{high throughput fractionated satellite}
\acrodef{HTS}{high throughput satellite}
\acrodef{UHF}{ultra-high frequency}
\acrodef{FF}{formation flying}
\acrodef{MIMO}{multiple input multiple output}
\acrodef{GNC}{guidance navigation and control}
\acrodef{NTN}{non-terrestrial network}
\acrodef{GNSS}{global navigation satellite system}
\acrodef{OFDM}{orthogonal frequency division multiplexing}
\acrodef{PHY}{physical-layer}
\acrodef{SA}{satellite array}
\acrodef{SADM}{solar array drive mechanism}
\acrodef{SG}{solar generator}
\acrodef{SNR}{signal-to-noise ratio}
\acrodef{SIR}{signal-to-interference ratio}
\acrodef{SINR}{signal-to-interference-plus-noise ratio}
\acrodef{SSPA}{solid-state power amplifier}
\acrodef{RF}{radio frequency}
\acrodef{R-GEO}{regional \acs{GEO}}
\acrodef{UT}{user terminal}
\acrodef{UE}{user equipment}
\acrodef{TDM}{time division multiplexing}
\acrodef{RRM}{radio resource management}
\acrodef{SSPA}{solid-state power amplifier}
\acrodef{BLER}{block error rate}
\acrodef{DL}{down-link}
\acrodef{UL}{up-link}
\acrodef{RV}{random variable}
\acrodef{PSD}{power spectral density}
\acrodef{UDSM}{ultra deep sub-micron}
\acrodef{PEPS}{power energy platform simulation}
\acrodef{SE}{spectral efficiency}
\acrodef{wrt}{with respect to}
\acrodef{OBP}{on-board digital processor}
\acrodef{UDSM}{ultra-deep-sub-micron}
\acrodef{UC}{user-centric}
\acrodef{CSI}{channel state information}
\acrodef{PAC}{per-antenna constraint}
\acrodef{FPAC}{fair \acs{PAC}}
\acrodef{MPC}{maximum power constraint}
\acrodef{AWGN}{additive white Gaussian noise}
\acrodef{TX}{transmit}
\acrodef{MMSE}{minimum mean square error}
\acrodef{MF}{matched filter}
\acrodef{ZF}{zero forcing}
\acrodef{ELSA}{enhanced logarithmic spiral array}
\acrodef{UPA}{uniform planar array}
\acrodef{NB}{narrowband}
\acrodef{WB}{wideband}
\acrodef{BFN}{beamforming network}
\acrodef{MD-MIQP}{minimum distance mixed integer quadratic problem}
\acrodef{PDF}{probability density function}
\acrodef{NPR}{noise-to-power ratio}
\acrodef{HPA}{high power amplifier}
\acrodef{CF}{cell-free}
\acrodef{3GPP}{third generation partnership project}
\acrodef{UC-MIMO}{user centric MIMO}
\acrodef{VHTS}{very high throughput satellites}
\acrodef{PM-MIMO}{pragmatic M-MIMO}
\acrodef{NR}{new radio}
\acrodef{LMS}{land mobile satellite}
\acrodef{IM}{intermodulation}
\acrodef{OBO}{output back-off}
\acrodef{FDD}{frequency division duplexing}
\acrodef{TDD}{time division duplexing}
\acrodef{BLER}{block error rate}
\acrodef{rv}{random variable}
\acrodef{COB}{center of beam}
\acrodef{ISL}{inter-satellite link}
\acrodef{AP}{access point}
\acrodef{SVD}{single value decomposition}
\acrodef{CW}{continuous wave}
\acrodef{AOCS}{attitude and orbit control system}
\acrodef{LSTM}{long short-term memory}
\acrodef{CPU}{central processing unit}
\acrodef{GPU}{graphics processing unit}
\acrodef{TPU}{tensor processing unit}
\acrodef{MMF}{max-min fairness}
\acrodef{MHA}{multi-head attention}
\acrodef{FFN}{feed-forward network}
\acrodef{MSE}{mean square error}
\acrodef{CDF}{cumulative distribution function}
\acrodef{EPA}{equal power allocation}
\acrodef{FPA}{fractional power allocation}
\acrodef{TNN}{transformer neural network}
\acrodef{RT}{real-time}
\acrodef{THz}{terahertz}
\acrodef{FNN}{feedforward neural network}
\acrodef{RNN}{recurrent neural network}
\acrodef{ULA}{uniform linear array}
\acrodef{UPA}{uniform planar array}
\acrodef{AoA}{angle of arrival}
\acrodef{AoD}{angle of departure}
\acrodef{NF}{near-field}
\acrodef{FF}{far-field}
\acrodef{ML}{machine learning}
\acrodef{LSTM}{long short-term memory}
\acrodef{DL}{deep learning}
\acrodef{CI}{close-in}

\begin{document}

\title{\LARGE{Deep Learning Prediction of Beam Coherence Time for Near-Field TeraHertz Networks}}
\author{Irched Chafaa\,\orcidlink{0000-0003-1467-5933}, 
            E. Veronica Belmega\,\orcidlink{0000-0003-4336-4704},~\IEEEmembership{Senior Member,~IEEE},
            Giacomo Bacci\,\orcidlink{0000-0003-1762-8024},~\IEEEmembership{Senior Member,~IEEE}
\vspace*{-0.5cm}
\thanks{This work was supported by the ``PHC Galil\'ee'' programme (project number: 52227RD), funded by the French Ministry for Europe and Foreign Affairs, the French Ministry for Higher Education and Research, the French-Italian University through the Galileo programme (project number G25-76), by the French National Research Agency (ANR-22-PEFT-0007) as part of France 2030 and the NF-FITNESS project, by the Italian Ministry of Education and Research (MUR) in the framework of the FoReLab Project (Departments of Excellence), 
and in part by the HORIZON-JU-SNS-2022 EU project TIMES under grant no. 101096307.
}
\thanks{I. Chafaa and G. Bacci are with the Dip. Ingegneria dell'Informazione, University of Pisa, 56122 Pisa, Italy. E.V Belmega is with Universit\'e Gustave Eiffel, CNRS, LIGM, F-77454, Marne-la-Vall\'ee, France and also with ETIS UMR 8051, CY Cergy Paris Universit\'e, ENSEA, CNRS, F-95000, Cergy, France. G. Bacci is also with Nat. Inter-University Cons. Telecommunications (CNIT), 43124 Parma, Italy. Emails: irched.chafaa@ing.unipi.it,  veronica.belmega@esiee.fr, giacomo.bacci@unipi.it.}
	}
\maketitle

\begin{abstract}
Large multiple antenna arrays coupled with accurate beamforming are essential in \ac{THz} communications to ensure link reliability. However, as the number of antennas increases, beam alignment (focusing) and beam tracking in mobile networks incur prohibitive overhead. Additionally, the near-field region expands both with the size of antenna arrays and the carrier frequency, calling for adjustments in the beamforming to account for spherical wavefront instead of the conventional planar wave assumption. In this letter, we introduce a novel \emph{beam coherence time} for mobile \ac{THz} networks, to drastically reduce the rate of beam updates. Then, we propose a deep learning model, relying on a simple feedforward neural network with a time-dependent input, to predict the beam coherence time and adjust the beamforming on the fly with minimal overhead. Our numerical results demonstrate the effectiveness of the proposed approach by enabling higher data rates while reducing the overhead, especially at high (i.e., vehicular) mobility.
\end{abstract}

\begin{IEEEkeywords}
Terahertz, deep learning, near-field, beam coherence time.
\end{IEEEkeywords}

\acresetall 

\vspace*{-0.5cm}

\section{Introduction}
\IEEEPARstart{T}{he} \ac{THz} band ($0.1-10\THz$) is considered a key enabler to meet the increasing demand for higher data rates and alleviate spectrum scarcity \cite{THzsurvey}. \ac{THz} communications employ large antenna arrays and beamforming techniques to combat the severe path loss, giving rise to two primary challenges. First, the large size of the arrays coupled with \ac{THz} frequencies lead to a significant expansion of the \ac{NF} region compared to sub-$6\GHz$ networks \cite{zhou2015spherical}. For example, the \ac{NF} region extends as far as $40$ meters from an \ac{AP} with $512$ antennas operating at $140\GHz$ \cite{cui2021near} and cannot be neglected. In other words, the system model used for beamforming design can no longer rely on the conventional planar wave assumption, but rather on the spherical wave model
\cite{zhou2015spherical}. Second, the beam of the \ac{AP} needs to remain constantly focused on the \ac{UE} to guarantee a reliable \ac{THz} link, leading to a critical beamforming overhead affecting the  performance (e.g., data rate, latency, etc.).

Hence, it becomes necessary to determine \emph{an adequate time interval between successive beam updates}, which takes into account the \ac{NF} region, to achieve a good tradeoff between beam steering overhead and communication performance. In \cite{theirTb, electronics, hur,khorsandmanesh2024beam}, a time duration, called the \emph{beam coherence time} $T_B$, is introduced for \ac{FF} mmWave networks. It is defined as the time during which the received signal power at the \ac{UE} remains consistently above a predefined threshold. Thus, the beam can be updated every interval of $T_B$ instead of every channel coherence time $T_C$. However, the proposed $T_B$ is determined by using approximate distributions of the beam gain pattern to represent the \ac{UE} received power. Furthermore, it relies solely on the \ac{FF} model, which makes it inapplicable for \ac{THz} \ac{NF} and hybrid \ac{NF}/\ac{FF} networks, given the nonnegligible range of the \ac{NF} region for large-array \ac{THz} systems. 

Inspired by the aforementioned works, in this letter we propose a novel beam coherence time tailored to \ac{NF} and hybrid \ac{NF}/\ac{FF} \ac{THz} communications. To this aim, we use the spherical wave assumptions to model channel characteristics and associated beamforming vectors, which generalizes the \ac{FF} model used in the literature. Unlike \cite{theirTb, electronics, hur}, we do not rely on approximations of the antenna gains distributions, but we rather exploit the accurate expressions for the beam gains to evaluate numerically the beam coherence time. Then, we propose a \ac{DL} model for predicting dynamically the beam coherence time, as it changes with the temporal variations of the wireless \ac{THz} channel, so as to enable an adaptive beamforming with little overhead in mobile networks. 

Our  learning model relies on a simple \ac{FNN} \cite{liu2017survey} that includes few (and fixed in the number) previously memorized parameters as inputs to account for the temporal dependencies and does not require specific and more complex architectures such as a \ac{RNN} \cite{liu2017survey}. The neural network training is done offline and relies on relevant simulated data. Two main advantages of our learning method are: 
\begin{enumerate*}[label=\emph{\roman*})]
\item low running complexity, as it allows to predict the beam coherence time faster than the numerical approach for the same system parameters and computation environment; required knowledge of the \ac{UE} position-related data (AP-UE distance, angle, speed), as opposed to the \ac{THz} channel state information required by the numerical method.
\end{enumerate*}

Numerical results show that our approach is compatible with \ac{NF} \ac{THz} networks and yields better data rate performance. Moreover, in highly mobile settings in which the overhead by updating the beam every $T_C$ is prohibitive (i.e., zero data rate), our method enables data transmission at a rate close to the ideal case, in which the beam overhead is neglected.  

\section{System Model and Problem Formulation}\label{sec:model}
\begin{figure}[t]
    \centering
  \includegraphics[width=0.9\columnwidth]{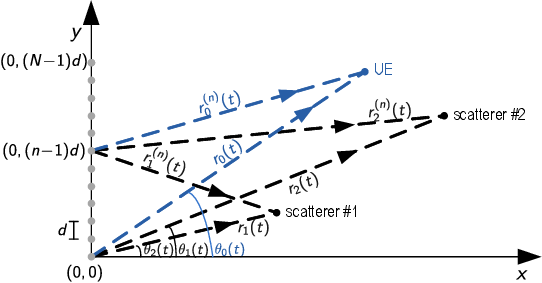}
   \caption{\acs{THz} link between a large array \acs{AP} and a \acs{UE}.}
    \label{fig:sys}
    \end{figure}

We consider a \ac{THz} communication link between an \ac{AP} and a mobile \ac{UE}, as shown in \figurename~\ref{fig:sys}. The \ac{AP} is equipped with a \ac{ULA} of $N$ elements, equispaced by distance $d$, whereas the \ac{UE} has a single antenna. This simple model allows us to focus on the essential problem of dynamically predicting the beam coherence time. However, the proposed model can be extended to a general framework by adapting the input features and the training dataset accordingly.
\subsection{Channel model}
To include the \ac{NF} effect, the spherical wave model is used to characterize the hybrid \ac{FF}/\ac{NF} channel at time $t$ \cite{busari2019generalized,wang2024fast}:
\begin{align}\label{eq:h}
    \mathbf{h}(t) = \sum_{\ell=0}^{L} g_\ell(t) e^{-j\frac{2\pi}{\lambda}r_\ell(t)} e^{j 2\pi f_{d,\ell} t} \mathbf{a}[\theta_\ell(t), r_\ell(t)],
\end{align} 
where $\lambda=c/f_c$ is the carrier wavelength, with $c$ and $f_c$ denoting the speed of light and the carrier frequency, respectively, whereas $L$ is the number of \ac{NLoS} paths, given by the presence of $L$ scatterers (in addition to the \ac{LoS} path, indexed by $\ell=0$ throughout the letter); $g_\ell(t)$ is the path gain of the $\ell$th path, $r_\ell(t)$ and $\theta_\ell(t)$ are the distance and the \ac{AoD} between the first \ac{AP} antenna element (also known as the reference antenna) and the \ac{UE} ($\ell=0$) or the scatterer ($\ell>0$), respectively; $f_{d,\ell}=\frac{v(t)}{\lambda}\cos\theta_\ell(t)$ is the Doppler frequency shift due to the radial velocity of the UE relative to the $\ell^{\text{th}}$ path, whereas $\mathbf{a}[\theta_\ell(t), r_\ell(t)]$ is the \ac{NF} steering vector, given as \cite{wang2024fast}:
\begin{align}\label{eq:sv}
  &\mathbf{a}[\theta_\ell(t), r_\ell(t)] \triangleq \nonumber\\
  &\quad\frac{1}{\sqrt{N}} 
\begin{bmatrix}
e^{-j\frac{2\pi}{\lambda} [r_\ell^{(1)}(t) - r_\ell(t)]}, \,
\ldots, \,
e^{-j\frac{2\pi}{\lambda} [r_\ell^{(N)}(t) - r_\ell(t)]}
\end{bmatrix},
\end{align}
where $r_\ell^{(n)}(t)$ denotes the distance between the $n$th \ac{ULA} element and the \ac{UE} ($\ell=0$) or the scatterer ($\ell>0$) (which implies $r_\ell^{(1)}(t)=r_\ell(t)$). Note that all quantities in \eqref{eq:h} show the dependence on time $t$, to account for the relative movements (especially in the case of the \ac{UE}). Following  the geometric relationships shown in \figurename~\ref{fig:sys}, we get
\begin{align}\label{eq:rn}
    r_\ell^{(n)}(t) \!=\! \sqrt{r^2_\ell(t) + (n\!-\!1)^2 d^2 - 
    2r^2_\ell (n\!-\!1) d \sin(\theta_\ell(t))}.
\end{align}

\subsection{Mobility model}\label{mobility}
We use the Gauss-Markov mobility model \cite{camp2002survey}, which updates the speed and direction of the \ac{UE}, at time $t+\delta$, based on their previous values at time $t$ as follows: 
\begin{align}
\begin{aligned}\label{eq:mobility}
v(t+\delta) &= \alpha\ v(t) + (1 - \alpha)\bar{v} + \sqrt{1 - \alpha^2} \  V, \\
\phi(t+\delta) &= \alpha\ \phi(t) + (1 - \alpha)\bar{\phi} + \sqrt{1 - \alpha^2} \ \Phi,
\end{aligned}
\end{align}
where $0 \leq \alpha \leq 1$ is a tuning parameter to control the randomness; $\bar{v}$  and $\bar{\phi}$ are average speed and direction, respectively; and $V$ and $\Phi$ are standard normal random variables. Note that all such parameters do not depend on $\delta$. However, as current values \eqref{eq:mobility} are correlated (through $\delta)$ with their previous ones, this mobility model provides smooth (and realistic) trajectory changes, reducing abrupt stops and turns. The \ac{UE}'s position at time $t$ is then updated as:
\begin{align}
\begin{aligned}
x(t+\delta) = x(t) + v(t) { \delta} \cos[\phi(t)], \\
y(t+\delta) = y(t) + v(t) { \delta} \sin[\phi(t)].
\end{aligned}   
\end{align}  
\subsection{Problem formulation}
To account for the \ac{THz} channel characteristics and the \ac{UE} mobility, the \ac{AP} needs to  update the beamforming vector $\mathbf{f}(t)=\mathbf{a}[\theta_\ell(t), r_\ell(t)]$, ideally every time the channel \eqref{eq:h} changes significantly. A fundamental tradeoff raises between  overhead cost and communication performance. On the one hand, we could consider updating the beam every channel coherence time $T_C$, resulting in high pointing accuracy, and hence high \ac{SNR}, at the cost of a significant overhead, which may significantly impact data transmission. On the other hand, less frequent beam updating reduces the overhead, but  degrades the \ac{UE} received \ac{SNR}. 

Suppose to maintain the same vector $\mathbf{f}(t)$ for a certain period: after a time $\tau$, the \ac{SNR} $\gamma(t+\tau)$ is
\begin{align}\label{eq:snr}
  \gamma(t+\tau)=\frac{P_T}{\sigma^2}
  \left|\mathbf{h}^H(t+\tau)\mathbf{f}(t)\right|^2\triangleq
  \frac{P_T}{\sigma^2} G(t+\tau)
\end{align}
where $P_T$ and $\sigma^2$ are the transmit and the \ac{AWGN} powers, respectively, and $G(t+\tau)$ is the beam gain at time $t+\tau$. 

As suggested in \cite{theirTb, electronics} for mmWave networks, a good tradeoff is to update the beam only when $G(t+\tau)$ (and hence the \ac{SNR}) falls below an acceptable threshold $\xi \in [0,1]$. This means defining a beam coherence time $T_B > T_C$ such that 
\begin{align}\label{Tbdef}
    T_B = \inf_{\tau}\left\{\tau, \frac{G(t+\tau)}{G(t)}\le\xi\right\}
\end{align}
In this paper, our main goal is to answer the following question: \emph{how can we determine a beam coherence time for \ac{THz} networks while taking into account the \ac{NF} wave propagation?} 

Deriving a closed-form expression of $T_B$ for a multi-path \ac{THz} channel, by solving analytically \eqref{Tbdef} under the assumption of perfect \ac{CSI} \eqref{eq:h} at each time $t+\tau$, is not straightforward. Instead, this can be done via exhaustive search. However, this approach shows two main shortcomings, making it unsuitable in dynamic and mobile \ac{THz} networks: 
\begin{enumerate*}[label=\emph{\roman*})]
\item required knowledge of the \ac{THz} channel, including time evolution of $\{\theta_\ell(t), r_\ell(t)\}_{l=0}^{L}$; and 
\item important computational cost of the numerical approach (e.g., with $N=512$ and $f_c=142\GHz$, it takes on average $2$ seconds).
\end{enumerate*} 
\section{Beam Coherence Time Prediction}
\begin{figure}[t]
    \centering
  \includegraphics[width=\columnwidth]{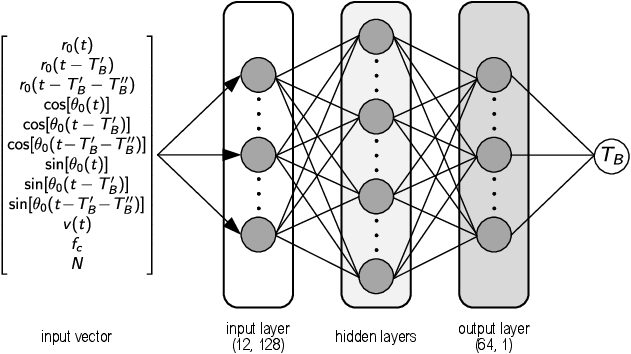}
   \caption{Diagram of the prediction model.}
    \label{FNN}
    \vspace*{-0.3cm}
\end{figure}

To overcome the issues of computing $T_B$ numerically, we propose here a \ac{DL} model to predict $T_B$. The numerical approach is used as a baseline to construct the labels for the offline training. Specifically, we consider the  \ac{FNN} architecture illustrated in \figurename~\ref{FNN}. The input features of the \ac{FNN} are composed of kinematic and system-level parameters relevant to beam dynamics, containing:
\begin{itemize}
\item the \ac{UE}'s speed $v(t)$, distance $r_0(t)$, and \ac{AoD} $\theta_0(t)$;
\item the carrier frequency $f_c$;
\item the number of \ac{AP} \ac{ULA} elements $N$.
\end{itemize}
To account for the temporal evolution of $\theta_0(t)$  and $r_0(t)$, we consider three consecutive snapshots, at times $t$, $(t-T_B^\prime)$, and $(t-T_B^{\prime}-T_B^{\prime\prime})$, where $T_B^\prime$ and $T_B^{\prime\prime}$ represent the two previous coherence beam times. The main motivation is as follows. On one side, to capture the dynamics of \ac{UE} mobility, we need to infer high-order motion characteristics, such as speed, acceleration, and changes of direction, which requires at least three successive time steps. On the other side, increasing the number of time steps increases complexity and computational cost. Furthermore, given the fast variability of \ac{THz} channels, considering too many consecutive time steps may provide an irrelevant history to $T_B$ estimation. Thus, three time steps offer a good tradeoff without resorting to more complex architectures, such as \ac{RNN} or \ac{LSTM} \cite{liu2017survey}.

This input requires an initialization phase at the beginning of deployment, when the first two values of $T_B$, obtained from the numerical solver, are needed to form the initial input vector. After the first two estimates, the model uses its own past predictions to construct the input, and  does not rely on the numerical solver anymore. Note also that, to preserve continuity and to account for periodicity of the \ac{AoD} $\theta_0(t)$, the \ac{FNN} is fed with both its sine and its cosine. In summary, the \ac{FNN} input is a $12$-dimensional vector (\figurename~\ref{FNN}),  capturing  the time evolution of \ac{UE} mobility (implicitly including radial velocity) and system  parameters that directly influence the predicted output $T_B$ at time $t$.
  
The proposed prediction model is trained to learn the unknown relation between the wireless environment dynamics and the corresponding beam coherence time. Once trained, the model can predict $T_B$ based on the \ac{UE} state (e.g., location, velocity, angle) and generalize over a wide range of mobility types and \ac{THz} frequencies. In  what follows, we provide the details about \ac{FNN} architecture and the dataset construction.

\subsection{Dataset construction}
First, several trajectories for different \ac{UE} mobility types are generated, using the Gauss-Markov model in \sectionname~\ref{mobility}. Then, $T_B$ is evaluated by solving numerically \eqref{Tbdef} for $f_c \in \{142,280\}\GHz$. The dataset, with $2.5 \times 10^5$ samples, is built by associating each $12$-dimension vector, which accounts for mobility features and system parameters, to the target output $T_B$. The input features of distance, angle and speed are contaminated with zero-mean Gaussian  noise, whose standard deviation is taken randomly in the range $[0,1]\meter$, $[0,5]\,^\circ$, and $[0,1]\mps$, respectively to make the model robust against input errors.
Finally, samples are normalized and split into training, validation and testing sets with the ratio $80\%$, $10\%$ and $10\%$. 
 
\subsection{\acs{FNN} architecture details}
The prediction model is a deep \ac{FNN} composed of $6$ fully connected layers with the following output dimensions: $128$, $256$, $512$, $256$, $64$, and $1$, respectively. Each hidden layer is followed by a LeakyReLU \cite{liu2017survey} activation function (with a negative slope of $0.01$), and a dropout layer with a dropout rate of $0.2$ to prevent overfitting, except the final output layer that uses a ReLU \cite{liu2017survey} activation to ensure nonnegative predictions. The model is trained using the AdamW optimizer \cite{liu2017survey} with an initial learning rate of $10^{-3}$ and a weight decay of $10^{-5}$. A learning rate scheduler is used to dynamically adjust the learning rate throughout training. The loss function  is a smooth variant of the mean absolute error (smooth L1 loss). The batch size is set to $64$, and the model is trained for $100$ epochs. 

This architecture is chosen due to its high capacity to model complex nonlinear relationships between the input and the output while keeping a low latency for the inference. For example, the measured average inference time per sample is only $5.3\mus$. The best model, based on the validation loss, is saved for evaluation on the test set, as illustrated below.

\section{Numerical Results}
We present below our numerical results to illustrate the performance of the proposed beam coherence time in a \ac{THz} link, using the model described in \sectionname~\ref{sec:model}. 

    \vspace*{-0.3cm}

\subsection{System parameters}
The system consists of a stationary \ac{AP} and a mobile \ac{UE}, located within the zone delimited by $x \in [0,50]\meter$ and $y \in [-25,25]\meter$. We also assume $L=2$ scatterers, randomly located in the same region. The \ac{AP} is equipped with $N=512$ antennas, with spacing $d=\lambda/2$, and using $P_T=30\dBm$ transmit power. The \ac{AP} incurs an overhead $T_\mathsf{ovh}=40\mus$, as estimated in \cite{shen2024energy}, for a single \ac{THz} beam, representing beam training, request, feedback, and acknowledgment phases. The carrier frequency and bandwidth are $f_c=142\GHz$ and $B=20\MHz$, respectively. The paths gains are generated using the \ac{CI} model in \cite[Eq. (1)]{xing2021propagation}  with a path loss exponent of $2.1$ (resp., $3.1$) and a large-scale shadow fading modeled as a zero-mean Gaussian random variable with a standard deviation of $2.8\dB$ (resp., $8.3\dB$) for the \ac{LoS} (resp., \ac{NLoS}) path. The noise power is $\sigma^2=-94\dBm$, with a noise figure $\eta=7\dB$. The threshold $\xi$ is set at $1/2$, which corresponds to a $3$-dB loss of the beam gain \cite{khorsandmanesh2024beam}. 

The \ac{UE} mobility follows the Gauss-Markov model described in \sectionname~\ref{mobility}, according to three categories. Each category is determined by parameters controlling the velocity and direction variations, which are defined to reflect realistic mobility across a range of urban scenarios. Specifically, \emph{pedestrians} have speeds uniformly distributed between $0.5\mps$ and $1.5\mps$, with a randomness coefficient $\alpha=0.3$ and a directional variation range of $\left[ \frac{\pi}{2}, \frac{3\pi}{4} \right]$, allowing for more fluctuation in movement direction. \emph{Bicycle} mobility is defined by speeds ranging from $2\mps$ to $6\mps$, a moderate memory factor of $\alpha=0.5$, and smoother direction transitions within the interval $\left[ \frac{\pi}{4}, \frac{\pi}{2} \right]$. \emph{Vehicle} mobility features velocities between $10\mps$ and $25\mps$, a strong temporal correlation coefficient $\alpha=0.7$, and less directional deviation within $\left[ 0, \frac{\pi}{8} \right]$.
 
\subsection{Beam coherence time prediction} 
The \ac{AP} updates its beam as follows: first, the beamforming vector $\mathbf{f}(t)$ is determined using the \ac{UE}'s current position $\left(\theta_0(t),r_0(t)\right)$. Then, the beam coherence time $T_B$ is \emph{predicted} by the learning model. During the life-time of the currently predicted $T_B$, the channel changes due to the \ac{UE} mobility, while the beam remains unchanged. At time $(t+T_B)$, a new beamforming vector $\mathbf{f}(t+T_B)$ and a new prediction of the coherence beam time are determined, using the updated \ac{UE}'s position $\left(\theta_0(t+T_B),r_0(t+T_B)\right)$. These steps are reiterated upon the expiration of the new coherence beam time. 

We compare the results of our approach, labeled \emph{predicted $T_B$}, with the following benchmarks:
\begin{itemize}
\item \emph{upper bound}: an ideal policy having access to an instantaneous and perfect \ac{CSI}, where the beam is updated every channel coherence time $T_C = \lambda/(4\bar{v})$ \cite{khan2012millimeter}, where $\bar{v}$ is the mean speed of the \ac{UE} mobility category, by ignoring any delay or overhead;
\item \emph{statistical $T_C$}: the beam is updated every $T_C$, computed as listed above, but taking into account the corresponding beamforming overhead;
\item \emph{numerical $T_B$}: the beam is updated every $T_B$, computed numerically solving \eqref{Tbdef}, which represents the ground-truth label of the testing data, taking into account the beamforming overhead.
\end{itemize}

  \begin{figure}[btp]
  \begin{center}
  \subfigure[\acs{UE} category: pedestrians]
    {\includegraphics[width=0.9\columnwidth]
    {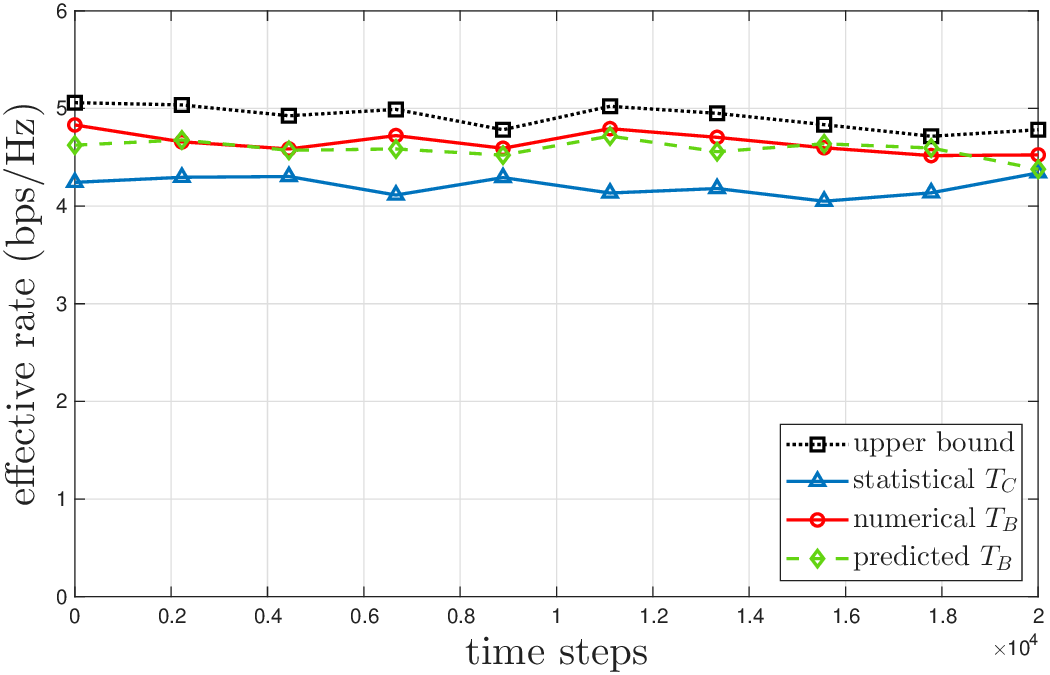}}
  \subfigure[\acs{UE} category: bicycles]
    {\includegraphics[width=0.9\columnwidth]
    {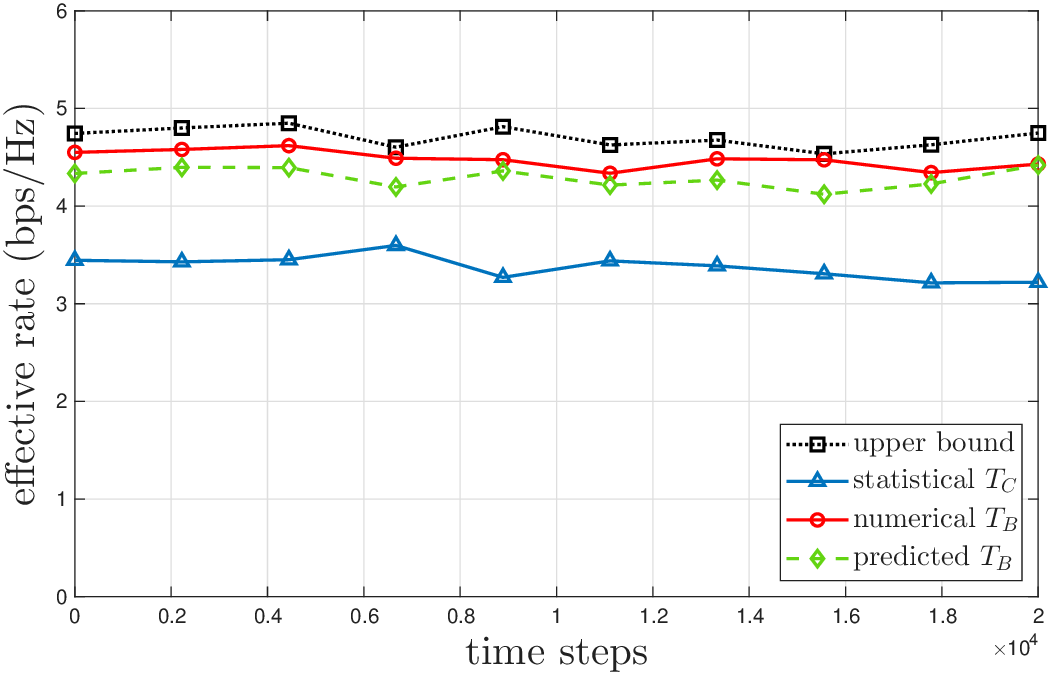}}
  \subfigure[ \acs{UE} category: vehicles]
    {\includegraphics[width=0.9\columnwidth]
    {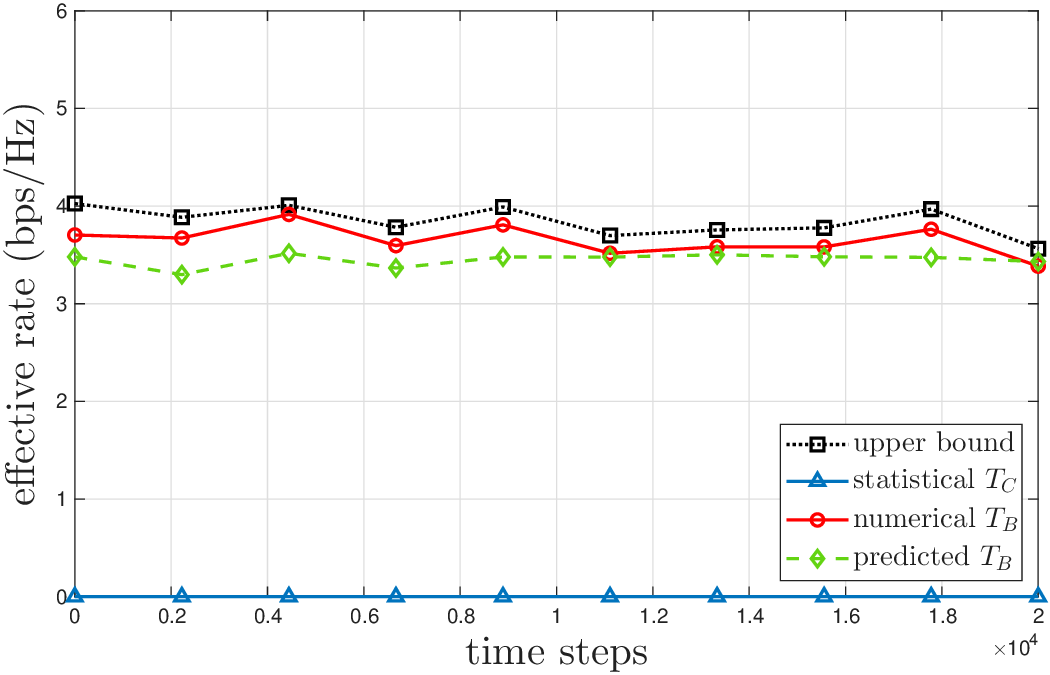}}
    \caption{Effective rate achieved using different beam durations for various mobility types. Updating the beam every $T_B$ yields a better tradeoff between rate performance and beam overhead.}
    \label{sim1}
  \end{center}
\vspace*{-0.5cm}
\end{figure} 

\paragraph{Effective rate performance} 

In \figurename~\ref{sim1}, we evaluate the temporal evolution of the effective data rate during a time window of $10$ seconds for three types of \ac{UE} mobility. The duration of each time step, which corresponds to the update time $\delta$ of our Gauss-Markov model, is $\delta=0.5\ms$. The results are averaged over $100$ trajectories for each type of mobility. The effective rate for each policy is evaluated using
\begin{align}\label{eq:rate}
  R_\mathsf{eff}(t+\delta;T)=
  \left[\left(1-\frac{T_\mathsf{ovh}}{T}\right)\log_2\left(1+\gamma(t+\delta)\right)\right]^+,
\end{align}
where $[x]^+=\max(0,x)$, the \ac{SNR} $\gamma(t+\delta)$ is defined as in \eqref{eq:snr}, and $T=T_C$ when using the policies for the upper bound and the statistical $T_C$, and $T=T_B$ when applying the policies based on $T_B$ (both the numerical and the predicted ones). Note that, when considering the upper bound, $T_\mathsf{ovh}=0$ in \eqref{eq:rate}.

Looking at \figurename~\ref{sim1}, we can see that dynamically adjusting the beam update time using $T_B$ enables near-optimal performance across different mobility regimes, while significantly reducing the overhead compared to frequent, every $T_C$, beam updates. Moreover, the effective rate achieved with the predicted $T_B$ closely follows the ones obtained by the ground-truth in all mobility scenarios. This highlights that the proposed prediction model obtains a high average rate performance while reducing the beamforming overhead. Since our learning model performs closely to the ideal upper-bound, we omit other baselines and limit the comparison to the exhaustive search method.

\paragraph{Impact of \acs{UE} mobility}
\figurename~\ref{sim1} also illustrates the effect of \ac{UE} mobility on $R_\mathsf{eff}$. For pedestrians, all policies perform similarly due to the low mobility of \acp{UE}. In the case of bicycles, updating the beam every $T_C$ starts to underperform as mobility increases compared to pedestrians. The $T_B$ approaches mitigate this loss effectively by adapting the update intervals to channel dynamics, yielding rates close to the upper bound. For vehicles (high mobility), the gap between $T_C$ and both $T_B$ approaches becomes remarkable. The $T_C$ policy suffers because of too frequent updates, resulting in \emph{no data} transmission. On the other hand, updating the beam every $T_B$ allows a significantly better tradeoff.

\begin{figure}[t]
    \centering
  \includegraphics[width=\columnwidth]{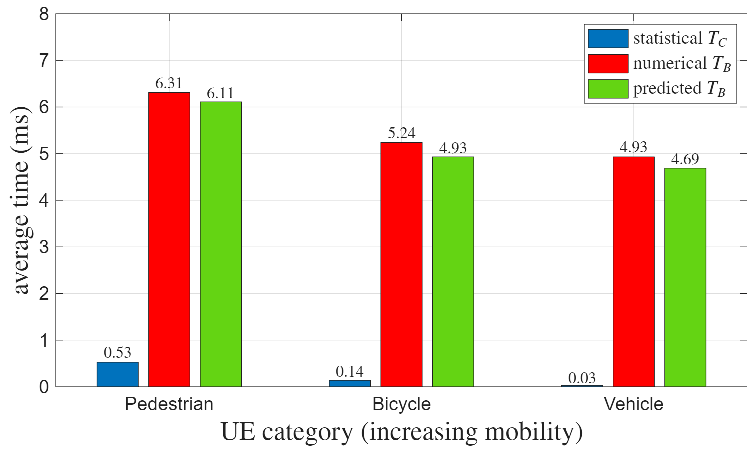}
   \caption{Average beam-update duration for different \acs{UE} mobility profiles. $T_B$ remains larger than $T_C$, yielding less overhead. When the speed increases, $T_B$ decreases as expected.}
    \label{sim2}
    \vspace*{-.5cm}
    \end{figure}

\paragraph{Beam duration} 
\figurename~\ref{sim2} shows the average beam-update durations. We can observe that $T_C$ is significantly shorter than $T_B$ (in both numerical and predicted approaches) for all mobility types. This confirms that relying on $T_C$ for beam updates results in excessively frequent updates, and consequently high, or even prohibitive, overhead. As the \ac{UE} speed increases, the average beam duration decreases, with the predicted $T_B$ adapting accordingly, which reflects the expected decrease in beam coherence time due to faster channel variations at higher speeds. Finally, the close numerical and predicted $T_B$ values  across all scenarios demonstrate the effectiveness of our \ac{DL} approach, which is able to capture the underlying channel time dynamics and adjust the beam duration accordingly.

\begin{figure}[t]
    \centering
  \includegraphics[width=0.9\columnwidth]{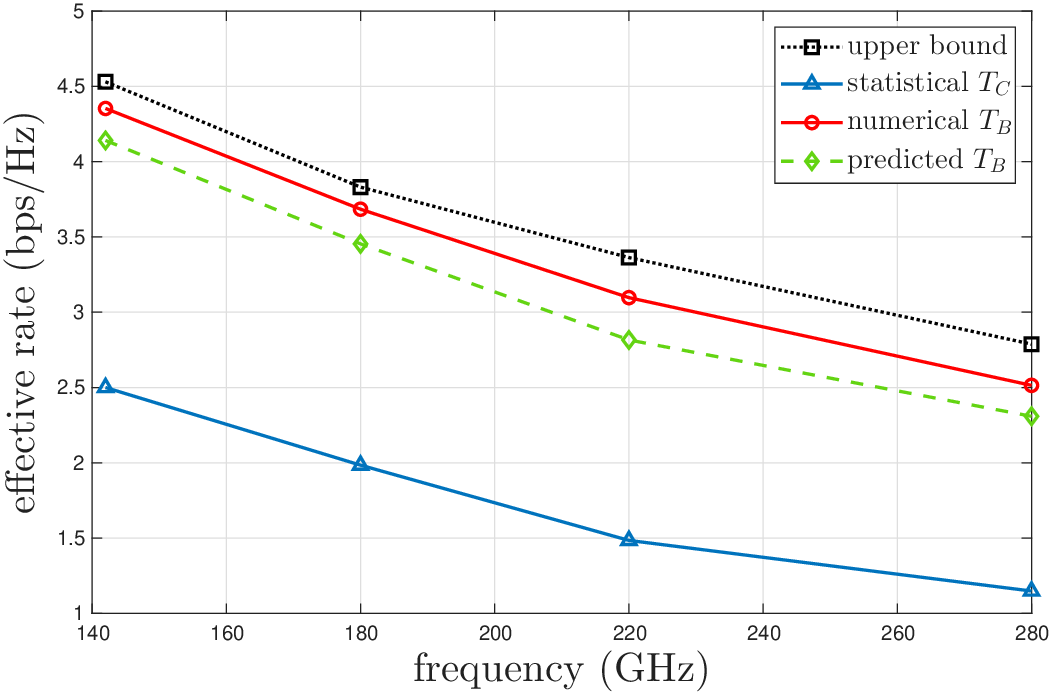}
   \caption{Average effective rate for different \acs{THz} frequencies $f_c$. The prediction model effectively adapts to $f_c$ variations.}
    \label{sim3}
    \vspace*{-.2cm}
    \end{figure}

\paragraph{Impact of carrier frequency} 
\figurename~\ref{sim3} evaluates the effect of increasing the \ac{THz} carrier frequency $f_c$ on $R_\mathsf{eff}$ for different strategies. Note that the plotted values are averaged across all mobility types. As $f_c$ increases from $142\GHz$ to $280\GHz$, all curves exhibit a decreasing $R_\mathsf{eff}$. This is primarily due to larger pathloss and smaller channel/beam life-times at higher frequencies, which increases the channel/beam update overhead. However, the results of our method remains close to the upper bound and the ground-truth $T_B$, showing that the prediction model effectively adapts the beam update time in response to frequency-dependent channel variations, even without explicit knowledge of $T_C$, and considering that the model is trained only for $142$ and $280\GHz$.

\section{Conclusion}
In this letter, we proposed a \ac{DL} model to predict the beam coherence time for \ac{NF} \ac{THz} communications with mobile \acp{UE}. Our approach considers the spherical wave model to characterize both the channel and beamforming vectors. The proposed learning model is a simple \ac{FNN} with a time-dependent input information, which allows  to handle \ac{UE} mobility and other temporal channel variations while reducing the prediction complexity. Numerical simulations show that our approach is better suited for \ac{NF} \ac{THz} networks with mobility in terms of effective rate and inference time. Our proposed method enables the \ac{AP} to assess an appropriate beam update rate in \ac{NF} systems and, hence, to reduce the beamforming overhead. Future work includes joint prediction of beam coherence time and an optimal beamforming vector in a multi-user scenario.



\bibliographystyle{IEEEtran}
\bibliography{IEEEabrv,biblio}

\begin{thebibliography}{10}
\providecommand{\url}[1]{#1}
\csname url@samestyle\endcsname
\providecommand{\newblock}{\relax}
\providecommand{\bibinfo}[2]{#2}
\providecommand{\BIBentrySTDinterwordspacing}{\spaceskip=0pt\relax}
\providecommand{\BIBentryALTinterwordstretchfactor}{4}
\providecommand{\BIBentryALTinterwordspacing}{\spaceskip=\fontdimen2\font plus
\BIBentryALTinterwordstretchfactor\fontdimen3\font minus
  \fontdimen4\font\relax}
\providecommand{\BIBforeignlanguage}[2]{{%
\expandafter\ifx\csname l@#1\endcsname\relax
\typeout{** WARNING: IEEEtran.bst: No hyphenation pattern has been}%
\typeout{** loaded for the language `#1'. Using the pattern for}%
\typeout{** the default language instead.}%
\else
\language=\csname l@#1\endcsname
\fi
#2}}
\providecommand{\BIBdecl}{\relax}
\BIBdecl

\bibitem{THzsurvey}
I.~F. Akyildiz, C.~Han, Z.~Hu, S.~Nie, and J.~M. Jornet, ``Tera{H}ertz band
  communication: An old problem revisited and research directions for the next
  decade,'' \emph{{IEEE} Trans. Commun.}, vol.~70, 2022.

\bibitem{zhou2015spherical}
Z.~Zhou, X.~Gao, J.~Fang, and Z.~Chen, ``Spherical wave channel and analysis
  for large linear array in {L}o{S} conditions,'' in \emph{Proc. IEEE Global
  Commun. Conf.}\hskip 1em plus 0.5em minus 0.4em\relax San Diego, CA: IEEE,
  2015.

\bibitem{cui2021near}
M.~Cui and L.~Dai, ``Near-field wideband beamforming for extremely large
  antenna arrays,'' \emph{{IEEE} Trans. Wireless Commun.}, vol.~23, 2024.

\bibitem{theirTb}
V.~Va, J.~Choi, and R.~W. Heath, ``The impact of beamwidth on temporal channel
  variation in vehicular channels and its implications,'' \emph{{IEEE} Trans.
  Veh. Technol.}, vol.~66, 2016.

\bibitem{electronics}
P.~Li, D.~Liu, X.~Hou, and J.~Wang, ``Trajectory prediction and channel
  monitoring aided fast beam tracking scheme at unlicensed mm{W}ave bands,''
  \emph{Electronics}, vol.~9, 2020.

\bibitem{hur}
S.~Hur, T.~Kim, D.~J. Love, J.~V. Krogmeier, T.~A. Thomas, and A.~Ghosh,
  ``Millimeter wave beamforming for wireless backhaul and access in small cell
  networks,'' \emph{{IEEE} Trans. Commun.}, vol.~61, 2013.

\bibitem{khorsandmanesh2024beam}
Y.~Khorsandmanesh, E.~Bj{\"o}rnson, J.~Jald{\'e}n, and B.~Lindoff, ``Beam
  coherence time analysis for mobile wideband {mmWave} point-to-point {MIMO}
  channels,'' \emph{{IEEE} Wireless Commun. Lett.}, vol.~13, 2024.

\bibitem{liu2017survey}
W.~Liu, Z.~Wang, X.~Liu, N.~Zeng, Y.~Liu, and F.~E. Alsaadi, ``A survey of deep
  neural network architectures and their applications,'' \emph{Neurocomputing},
  vol. 234, 2017.

\bibitem{busari2019generalized}
S.~A. Busari, K.~M.~S. Huq, S.~Mumtaz, J.~Rodriguez, Y.~Fang, D.~C. Sicker,
  S.~Al-Rubaye, and A.~Tsourdos, ``Generalized hybrid beamforming for vehicular
  connectivity using {THz} massive {MIMO},'' \emph{{IEEE} Trans. Veh.
  Technol.}, vol.~68, 2019.

\bibitem{wang2024fast}
H.~Wang, J.~Fang, H.~Duan, and H.~Li, ``Fast hybrid far/near-field beam
  training for extremely large-scale millimeter wave/terahertz systems,''
  \emph{{IEEE} Trans. Commun.}, vol.~73, 2025.

\bibitem{camp2002survey}
T.~Camp, J.~Boleng, and V.~Davies, ``A survey of mobility models for ad hoc
  network research,'' \emph{Wireless Commun. Mobile Comput.}, vol.~2, 2002.

\bibitem{shen2024energy}
L.-H. Shen, K.-T. Feng, and L.-L. Yang, ``Energy efficient beamforming training
  in terahertz communication systems,'' \emph{{IEEE} Trans. Veh. Technol.},
  vol.~74, 2024.

\bibitem{xing2021propagation}
Y.~Xing and T.~S. Rappaport, ``Propagation measurements and path loss models
  for sub-thz in urban microcells,'' in \emph{ICC 2021-IEEE International
  Conference on Communications}.\hskip 1em plus 0.5em minus 0.4em\relax IEEE,
  2021, pp. 1--6.

\bibitem{khan2012millimeter}
F.~Khan, Z.~Pi, and S.~Rajagopal, ``Millimeter-wave mobile broadband with large
  scale spatial processing for 5{G} mobile communication,'' in \emph{Proc.
  Allerton Conf. Commun., Control, Comp.}, Monticello, IL, 2012.

\end{thebibliography}


 




\end{document}